\documentclass[final,5p,times,twocolumn]{elsarticle}

\usepackage{graphicx}
\usepackage{amssymb}
\usepackage{amsmath}
\usepackage{booktabs}
\usepackage{algpseudocode}
\usepackage{amsmath}
\biboptions{square, comma, numbers, sort&compress}

\usepackage[usenames,dvipsnames]{color}
\definecolor{darkblue}{rgb}{0,0.08,0.45}
\usepackage[bookmarksopen=true,bookmarksopenlevel=2,colorlinks=true,linkcolor=darkblue,anchorcolor=darkblue,citecolor=darkblue,urlcolor=darkblue,pdfstartview=FitH]{hyperref}
\usepackage[all]{hypcap}
\usepackage{algorithm}

\newcommand{\var}[1]{\mathtt{#1}}
\newcommand{\func}[1]{\mathbf{#1}}
\newcommand{\set}[2]{#1 \leftarrow #2}
\newcommand{\load}[2]{#1 \Leftarrow #2}
\newcommand{\store}[2]{#1 \Rightarrow #2}
\journal{Journal of Computational Physics}

\newenvironment{DIFnomarkup}{}{}

\begin{document}
\begin{frontmatter}

\title{Massively parallel Monte Carlo for many-particle simulations on GPUs}

\author[che]{Joshua A. Anderson}
\author[che]{Eric Jankowski}
\author[mse]{Thomas L. Grubb}
\author[che]{Michael Engel}
\author[che,mse]{Sharon C. Glotzer}

\address[che]{Department of Chemical Engineering, University of Michigan, Ann Arbor, MI 48109, USA}
\address[mse]{Department of Materials Science and Engineering, University of Michigan, Ann Arbor, MI 48109, USA}

\begin{abstract}
Current trends in parallel processors call for the design of efficient massively parallel algorithms for scientific computing.
Parallel algorithms for Monte Carlo simulations of thermodynamic ensembles of particles have received little attention because of the inherent serial nature of the statistical sampling.
In this paper, we present a massively parallel method that obeys detailed balance and implement it for a system of hard disks on the GPU.
We reproduce results of serial high-precision Monte Carlo runs to verify the method.
This is a good test case because the hard disk equation of state over the range where the liquid transforms into the solid is particularly sensitive to small deviations away from the balance conditions.
On a Tesla K20, our GPU implementation executes over one billion trial moves per second, which is 148 times faster than on a single Intel Xeon E5540 CPU core, enables 27 times better performance per dollar, and cuts energy usage by a factor of 13.
With this improved performance we are able to calculate the equation of state for systems of up to one million hard disks.
These large system sizes are required in order to probe the nature of the melting transition, which has been debated for the last forty years.
In this paper we present the details of our computational method, and discuss  the thermodynamics of hard disks  separately in a companion paper.
\end{abstract}

\end{frontmatter}



\section{Introduction}

During the last decades computational scientists have enjoyed a doubling of performance for single-threaded applications every two years solely from improvements in computer architecture.
This is no longer the case.
Current processor designs run into the Power Wall, which limits attainable clock speeds in a given power budget, and the Instruction Level Parallelism (ILP) Wall, which exists because there is only so much ILP that can be extracted from a typical program~\cite{Asanovic2006}.
Moore's law still holds, for now, and the additional transistors go into increasing the core counts on each new chip.
Consequently, researchers must utilize parallelism to execute larger, longer, or more demanding calculations and simulations.

As of this publication, a cluster of networked multi-core CPUs is the most common system architecture.
In an alternative approach, a single graphics processing unit (GPU) can execute thousands of instructions at the same time and provides the performance of a small cluster at a fraction of the cost~\cite{Stone2010}.
GPUs are becoming popular as desktop `personal supercomputers' and as coprocessors in heterogeneous clusters.
A successful GPU algorithm divides a given computation into a maximal number of identical, fully independent, and simple tasks, called \emph{threads}.
To fully utilize their potential it is necessary to design not just parallel, but \emph{massively parallel} algorithms that scale to thousands of  threads.
Over the last few years, many problems have successfully been adapted to GPUs.
One example is the molecular dynamics (MD) method for simulating thermodynamic ensembles of particles, which is well suited for massive parallelization.
Numerous MD software packages support GPUs, including HOOMD-blue~\cite{Anderson2008a, HOOMDWeb}, LAMMPS~\cite{Brown2011}, AMBER~\cite{Gotz2012,Grand2012a}, NAMD~\cite{Stone2007}, OpenMM~\cite{Eastman2010}, FEN ZI~\cite{Ganesan2011}, HALMD~\cite{Colberg2011}, the work by Rappaport~\cite{Rapaport2011}, GROMACS~\cite{GROMACS} and ACEMD~\cite{ACEMD}.

A case where implementation to GPUs has so far not been achieved is Monte Carlo (MC) applied to off-lattice many-particle systems.
MC is a statistical, rather than deterministic, sampling method that, appropriately implemented, samples the microstates of desired thermodynamic ensembles.
It is the method of choice in many situations because it only requires an interaction potential, not a force field, allowing \textit{e.g.} the use of non-differentiable pair potentials.
Such potentials are useful in the simulation of hard particles, which interact solely via excluded volume. 
MC is also flexible in the sense that a wide variety of update moves can be applied~\cite{Swendsen1987,Liu2004,Whitelam2007,Bernard2009}.
MC is easy to implement on serial machines where each step selects a new microstate at random and accepts or rejects the new microstate based on the Boltzmann factor.
An example trial move involves translating a single particle in a random direction.
Since the acceptance of a trial move depends on results of prior moves, subsequent moves usually cannot be performed independently.
A massively parallel algorithm must not only update a large number of particles at the same time, but also do so in a statistically correct way.
This can be achieved by obeying detailed balance, which is not a trivial task.
For these reasons, parallel MC codes for particle systems have received much less attention in the literature than parallel MD.

Available parallel MC algorithms fall into several categories.
Most of them rely on domain decomposition schemes to update portions of the problem in parallel.
Lattice MC has been employed for Ising models~\cite{Pawley85, Ren2006} including GPU implementations~\cite{Preis2009, Levy2010}.
Related schemes for particle systems exist~\cite{Heffelfinger1996,Uhlherr2002, Ren2007, OKeeffe2009, Sadigh12}, but none scale up to thousands of threads.
Moreover, the updating of domains employed in some of those methods can introduce a sampling bias that precludes the balance conditions.
Asynchronous parallel algorithms~\cite{Lubachevsky1987, Korniss1999} are poorly suited for GPUs because of their extensive inter-thread communication.
Much work has been done on parallelizing Kinetic Monte Carlo simulations \cite{Martinez2008, Arampatzis2012}, but these do not directly apply to the systems we are interested in.
Hybrid approaches that employ MD trajectories to create trial configurations~\cite{esselink95, loyens95} can be an effective way to exploit GPU parallelism, but require substantial additional programming and are not guaranteed to evolve any faster than MD alone.

\begin{figure}[tb]
\centering
\includegraphics{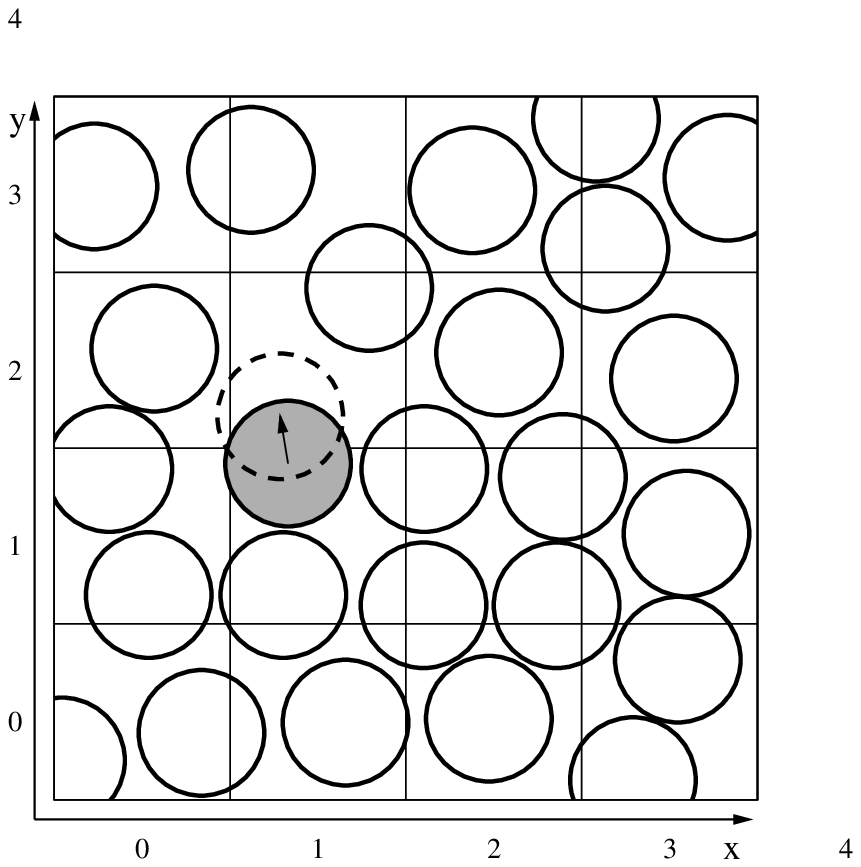}
\caption{\label{fig:spmc} In a serial hard disk simulation, each trial move typically consists of the following:
(i)~Select a single disk at random, (ii)~apply a random displacement to it, and (iii)~accept the move if it generates no overlaps.
The cell list enables $O(1)$ overlap checks by limiting the search space to only nine local cells.
A sweep is defined as $N$ consecutive trial moves, where $N$ is the number of disks in the simulation.}
\end{figure}

In this paper, we develop an algorithm for massively parallel Monte Carlo (MPMC) simulations of many-particle systems that obeys detailed balance.
As a test case, we implement it on the GPU for a system of hard disks in two dimensions (\autoref{fig:spmc}) and validate it with comparisons to recent large-scale serial event-chain MC simulations~\cite{Bernard2011}.
Our algorithm is not specific to this system, and it is valid for any MC simulation with local interactions between particles or lattice sites on any massively parallel computer architecture.

\section{Algorithm}

In developing any parallel application, the programmer must identify the computations that can be executed simultaneously.
As the total work is broken into smaller tasks, opportunities for scaling to more processors increase.
GPUs execute especially fine grained work loads.
In GPU MD applications, one thread typically acts on a single particle~\cite{Anderson2008a,Brown2011,Colberg2011,Stone2007,Rapaport2011}.
Such a decomposition is not directly applicable to traditional MC because each trial move depends on the state of the neighboring particles.

In MPMC, we utilize the \emph{cell list} data structure for parallel decomposition, as well as overlap checks in the case of hard particles.
A checkerboard decomposition permits many cells to be updated independently.~\cite{Pawley85} 
Although similar to applying trial moves to particles in a particular sequential order, checkerboard decomposition differs from the serial algorithm in one key way.
Particle positions, not labels (or indices), determine the order of updates, so the order will change as particles migrate.
Consequently, careless choices can lead to erroneous simulations.
We prove that our implementation of MPMC obeys detailed balance to ensure that no incorrect choices are made in its design.

\subsection{Checkerboard decomposition}
\label{sec:checker}

\begin{figure*}
\centering
\includegraphics{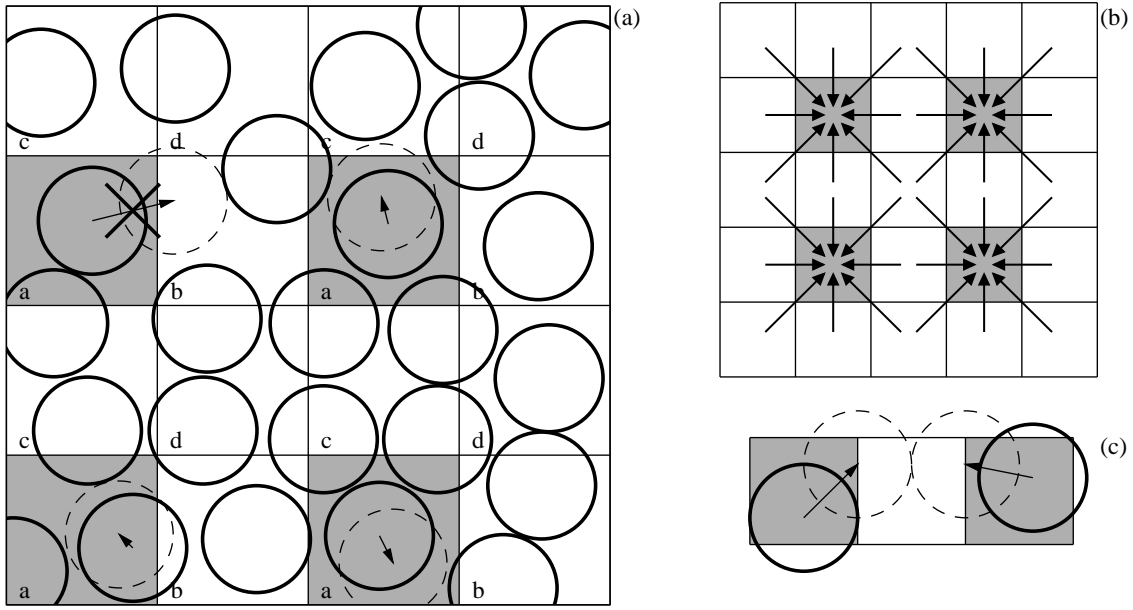}
\caption{\label{fig:mppmc} (a) In massively parallel Monte Carlo (MPMC), trial moves are concurrently applied to particles in a subset of the cells. Moves that leave the cell are rejected.
(b) Selected cells are separated by one row or one column of inactive cells.
During the evaluation of the acceptance criterion, each active cell reads the particles in the eight neighboring inactive cells.
(c) Simultaneous trial moves do not interact when the cell width is greater than the interaction range $\sigma$.}
\end{figure*}

The checkerboard domain decomposition scheme~\cite{Pawley85,Heffelfinger1996} divides the simulation volume into sets of square (cubic) cells (see \autoref{fig:mppmc}).
Checkerboarding maps well to MC simulations because it allows parallel updates of each set, comprising one quarter (one eighth in three dimensions) of the simulation volume.
The $x$ and $y$ coordinates of the cell (and $z$ in three dimensions) determine its checkerboard set $Q\in\{a,b,c,d,\ldots\}$:
\begin{equation}
Q = \left\{
\begin{array}{rlcl}
a & \text{if } (x\in \text{Even}) & \text{ and } & (y\in \text{Even}),\\
b & \text{if } (x\in \text{Odd}) & \text{ and } & (y\in \text{Even}),\\
c & \text{if } (x\in \text{Even}) & \text{ and } & (y\in \text{Odd}),\\
d & \text{if } (x\in \text{Odd}) & \text{ and } & (y\in \text{Odd}),\\
\ldots & \ldots,
\end{array} \right.
\label{checker}
\end{equation}
where $a,b,\ldots$ indicate labels of checkerboard sets.

The width of the cell $w$ must be chosen greater than the diameter of the disk $\sigma$ (generally, the pair interaction cutoff).
At the minimum $w=\sigma$, two particles separated by one cell can move without interacting (see \autoref{fig:mppmc}(c)).
Thus, the moves available to particles in a cell are independent from those in other cells of the same checkerboard set.

Most previous parallel MC simulations with mobile particles use stripe domain decomposition~\cite{Ren2007, Uhlherr2002, OKeeffe2009}, a one-dimensional version of the checkerboard decomposition, which minimizes the interface (and therefore communication) between domains.
However, the number of stripes and therefore the number of trial moves that can be conducted in parallel is low.
This means stripe decomposition is not efficient for parallelization on more than a few cores.

\subsection{Sweep structure}
\label{sec:sweep_structure}

\begin{algorithm}[tb]
\caption{Monte Carlo sweep}
\label{alg:mppmc}
\begin{algorithmic}[1]
\State $\set{C}{\{a,b,c,d, \ldots\}}$
\State $\func{rng.shuffle}(C)$
\For {$Q\in C$ } \Comment {Loop over sub-sweeps}
	\For {$c \in \mathrm{cells}(Q)$, \emph{in parallel}} \Comment{Loop over cells}
		\State $\func{rng.shuffle}(c.particles)$
		\For {$s \in [0 \ldots n_M)$}
		    \State $\set{p}{c.particles[\func{mod}(s,\func{len}(c.particles))]}$
			\State Generate trial move
			\If {$p$ remains in cell \emph{and} move accepted}
				\State Move $p$
			\EndIf
		\EndFor
	\EndFor
\EndFor

\State $\set{d}{\func{rng.uniform}(0,w/2)}$
\State $\set{\vec{f}}{\func{rng.choose}(-x, +x, -y, +y, \ldots)}$
\State $\func{shift\_cells}(\vec{f},d)$

\end{algorithmic}
\end{algorithm}

\autoref{alg:mppmc} outlines the structure of MPMC.
It splits each sweep over cells into sub-sweeps (four in two dimensions, eight in three dimensions), one handling each checkerboard set.
Line 2 shuffles the order of checkerboard sets using Fisher-Yates~\cite{Durstenfeld1964} to guarantee a random permutation.
During a sub-sweep, the algorithm concurrently processes all of the cells in the active set (line 4).

Each concurrent cell update shuffles and then loops over $n_M$ trial moves (lines 5,6).
Line 7 selects the particle from the cell, repeating from the start of the list when $n_M > n$, where $n$ is the number of particles in the cell.
Fixing the number of moves in all cells to the same number distributes computational effort most evenly among GPU cores~\cite{Krauth2012}.
Line 8 generates a trial move for each particle.
Line 9 accepts the move if it passes the normal Metropolis acceptance criterion~\cite{Metropolis1953} and the particle center remains in the cell~\cite{Uhlherr2002}.
Lines 15-17 maintain ergodicity by performing a cell shift, which redraws the cell boundaries in a randomly chosen location.

\subsection{Detailed balance}

A MC simulation obeys detailed balance if, for every internal process evolving the system there exists a reverse process occurring at the same rate.  
This ensures that a sequence of configurations converges to the correct equilibrium distribution, regardless of the initial condition.
Markov-chain MC generates a sequence of configurations where the probability $x_j(t+1)$ of observing the system in state $j$ at step $t+1$ is determined only by the previous state $i$ at step $t$.
This can be expressed by
\begin{equation}
x_j(t+1) = x_i(t)P_{ij},
\end{equation}
where $\boldsymbol{x}(t)=\{ x_1(t), x_2(t), ... , x_n(t) \}$ is the probability distribution at step $t$.
The elements $P_{ij}$ of the transition matrix represent the probabilities that the system will transition from state $i$ to state $j$.
If there exists an equilibrium distribution of states $\boldsymbol{x}^*$ for which $\boldsymbol{x}^*=\boldsymbol{x}^*P$, then $\boldsymbol{x}(t)$ is guaranteed to converge to $\boldsymbol{x}^*$ as $t\rightarrow \infty$ when $P$ satisfies detailed balance:
\begin{equation}\label{eqn:db}
x^*_i P_{ij} = x^*_j P_{ji}.
\end{equation}

Although detailed balance of a Markov chain is a sufficient condition to ensure convergence, it is not necessary.
Manousiouthakis and Deem show that an irreducible transition matrix that enforces regular sampling ($\exists m :  (P^m)_{ij}>0\, \forall i,j$) and obeys balance ($\boldsymbol{x^*}=\boldsymbol{x^*}P$) is both necessary and sufficient for convergence to the correct equilibrium distribution~\cite{deem99}.  
In this work we choose to enforce detailed balance.

The MPMC algorithm constructs a Markov chain and obeys detailed balance on the level of a MC sweep.
This follows directly from the observation that for each sweep there is exactly one inverse sweep, which can be seen as follows.
Take a particular sequence of sub-sweeps and a sequence of $n_M$ moves (either accepted or rejected) within each cell.
The reverse sweep consists of the reverse sequence of sub-sweeps and the reverse sequence of moves within each cell, with each move following the negative of the original vector.
For example, with $n_M=6$ trial moves per cell and $n=4$ particles in the cell, the original sequence would be $[0,1,2,3,0,1]$.
There is exactly one particle shuffling, $[1,0,3,2]$, that generates the reverse sequence $[1,0,3,2,1,0]$.
Since each sequence is chosen randomly from all possible permutations, the forward and reverse sequences occur with equal probability, and thus detailed balance holds.

\subsection{Pitfalls leading to incorrect statistical sampling}
Detailed balance is ensured when the following three steps are in place.
\begin{enumerate}
\item \emph{The particle center must not leave the cell}.~\cite{Uhlherr2002}
If particles are allowed to leave their cells during a sub-sweep, the reverse sequence of moves cannot be generated and detailed balance is not ensured.
When we skip this restriction in the hard disk system, it always develops order in the same orientation and at a lower than expected density.
\item \emph{Shuffling the particles in each cell.}
Particles entering a cell are added at the end of the cell list and cell lists are partially maintained during a cell shift.
When we skip particle shuffling, a temporal memory of previous states builds up over many sweeps, violating the Markov property.
\item \emph{Shuffling the checkerboard set.}
Without shuffling of the checkerboard set, the reverse sweep cannot be generated, which violates the condition of detailed balance.  
\end{enumerate}

To increase the number of accepted moves per sweep one might be tempted to allow particles to leave the cell, compensating the violation of step (1) by ensuring that each particle moves exactly once per sweep.
However, this procedure does not guarantee balance.
Cell updates with moves that leave a cell in the `middle' of the cell update (\textit{i.e.}~not the first or last successful move of the sub-sweep into a given neighboring cell) are not reversible and generate an incorrect probability distribution.
When we apply this scheme in the hard disk system, the pressure is shifted slightly away from the correct value and the magnitude of shift depends on the maximum trial move distance.

The particle shuffling step (2) is often explicitly omitted in favor of sequential updating~\cite{Ren2006, Ren2007, OKeeffe2009, Levy2010}.
As a justification, these authors refer to the analysis of Manousiouthakis and Deem~\cite{deem99}, who showed that shuffling is not necessary for the Ising lattice model away from infinite temperature.
However, their analysis cannot be transferred to systems of mobile particles if the sequence of particles is determined dynamically.
Our simulations for hard disks confirm (\autoref{tab:balancetest}) that skipping the particle shuffling step alters the pressure close to the melting transition.
In contrast, the checkerboard set shuffling step (3) is not necessary for correct sampling.

\begin{table}
\centering
\begin{tabular}{r r l l l}
\toprule
\multicolumn{2}{c}{Shuffling} & \multicolumn{3}{c}{$P^*$ in the hard disk system at} \\
P. & CB. & $\phi=0.698$ & $\phi=0.708$  & $\phi=0.716$ \\
\cmidrule(r){1-2}
\cmidrule(l){3-5}
Yes & Yes & 9.17079(5) & 9.18214(6) & 9.1774(2) \\
Yes & No  & 9.1707(1) & 9.1821(2) & 9.1775(3) \\
No & Yes  & 9.1716(1) & 9.1876(2) & 9.1831(3) \\
No & No  & 9.1715(1) & 9.1873(1) & 9.1823(2) \\
\bottomrule
\end{tabular}
\caption{\label{tab:balancetest}
Simulations of $N=256^2$ hard disks with particle (P.) and checkerboard (CB.) shuffling enabled or disabled.
The comparison of equilibrium pressures $P^*$ for runs at three different packing fractions $\phi$ demonstrates the necessity to shuffle particles.
Checkerboard shuffling is not required to obtain correct results.}
\end{table}


\section{Implementation}

We implement MPMC for hard disks using the NVIDIA CUDA programming model and execute benchmarks on a Tesla K20 graphics processor.
CUDA is an established parallel programming language; details may be found in the CUDA programming guide~\cite{NVIDIA2012}, text books~\cite{Kirk2010,Farber2011,Sanders2010} or in other publications, Refs.~\cite{Anderson2008a, Stone2010} for example.
The pseudocode presented in this paper is general enough that it could be adapted to any data-parallel language (\textit{e.g.}~OpenCL or OpenMP).

\subsection{Data Structures}

The proper choice of data structures can make or break an implementation's performance.
We keep all of the architecture details of NVIDIA GPUs in mind when designing our implementation.
MD codes store particles in a flat array with $N$ elements, and auxiliary data structures indirectly reference this list by index~\cite{Anderson2008a,Brown2011,Colberg2011,Rapaport2011}.
Such a data structure is not appropriate for MC as it would be expensive to rebuild the cell list every sweep.

Instead, we store the particle positions directly in the cell list.
That data is a sparse flat array, $\var{disk}[x,y,i]$, with storage for $m \cdot m \cdot n_\mathrm{max}$ particle positions, where $m$ is the number of cells on the side of the simulation box and $n_\mathrm{max}$ is the maximum number of particles allowed in a cell.
The auxiliary array, $\var{n}[x,y]$ stores the number of particles in each cell, where the particles are placed in elements $i \in [0 \ldots n)$.

\begin{figure}
\centering
\includegraphics{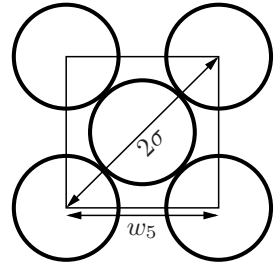}
\caption{\label{fig:maxcell} Four disks of diameter $\sigma$ are placed on the corners of the cell and a fifth in the center.
The smallest cell that can contain five disk centers has a diagonal of $2\sigma$ and an edge length of $w_\mathrm{5} = \sqrt{2}\sigma$.
Thus, the largest cell that can contain a maximum of four disk centers has a width $w < \sqrt{2}\sigma.$}
\end{figure}

We minimize the number of overlap checks and maximize parallelism by setting the cell width $w$ small, but not so small that a large fraction of moves will cross the cell boundaries.
The size must also be chosen so that $n_\mathrm{max}$ is known.
\autoref{fig:maxcell} shows that the largest cell that can fit no more than four particles has a width $w < \sqrt{2}\sigma$.
Furthermore, we avoid expensive boundary condition checks by choosing $m$ as a multiple of $2$ times the block size, because each thread handles every other cell.
For the 32-thread blocks used here, we set $m = \lfloor L / (\sqrt{2}\sigma) \rfloor$ and then round up to the nearest multiple of 64.

Each cell has a local coordinate system to mitigate floating point cancellation errors that would otherwise occur for large absolute coordinate values.
Whenever difference vectors are computed a coordinate system transformation is needed.
The components of the translation vector are either $+w$, $-w$, or 0 depending on the relative location of the neighboring cell.

\subsection{Kernel}

\begin{algorithm}
\begin{algorithmic}[1]
\caption{Sub-sweep GPU kernel}
\label{alg:subsweep}
\Require \texttt{gdim} is $(m / \var{bdim.x} / 2, m / \var{bdim.y} / 2)$
\Require $(\var{off.x},\var{off.y})$ is the offset to the lower-leftmost active cell in the sub sweep
\Require \texttt{seed} is a random number seed chosen by the user and fixed for the duration of a run
\Require \texttt{sweep} is the index of the current sweep
\Require $\vec{D}_\mathrm{sh}[p]$ is stored in shared memory such that each thread indexes unique elements in memory for $p \in [0 \ldots n_\mathrm{max})$

\State $\set{x}{2 \left(\var{bidx.x}\cdot \var{bdim.x} + \var{tidx.x}\right) + \var{off.x}}$
\State $\set{y}{2 \left(\var{bidx.y}\cdot \var{bdim.y} + \var{tidx.y}\right) + \var{off.y}}$

\State $\set{\var{rng}}{\func{Saru}(mx+y, \var{step}, \var{seed})}$
\State $\load{n}{\var{n}[x,y]}$

\If {$n == 0$}
	\State $\func{return}$
\EndIf

\State $\load{\vec{D}_\mathrm{sh}[i]}{\var{disk}[x,y,i]}~\forall i \in [0 \ldots n_\mathrm{max})$

\State $\func{rng.shuffle}(\vec{D}_\mathrm{sh}[0 \ldots n])$
\State $\set{i}{0}$
\For {$s \in [0 \ldots n_M)$}
	\State $\set{ \vec{D}_\mathrm{move} }{ \vec{D}_\mathrm{sh}[i] +  \func{rng.inCircle}(d)}$
	\State $\set{ \var{overlap} }{ \mathbf{False} }$
	\For{ $(x_\mathrm{neigh}, y_\mathrm{neigh}) \in$ neighborhood of cell $(x,y)$ }
		\State $\set{ \vec{s} }{ $vector pointing to current neighbor }
		\For{ $j \in [0 \ldots n_\mathrm{max}) $  }
			\State \textbf{continue} when ($x_\mathrm{neigh}, y_\mathrm{neigh})= (x,y)  \wedge i = j$
			\State $\load{ \vec{D} }{ \var{disk}[x_\mathrm{neigh}, y_\mathrm{neigh},j] }$
			\If { $\left| \vec{D}_\mathrm{move} - (\vec{D} + \vec{s}) \right| < \sigma$ }
				\State $\set{ \var{overlap} }{ \mathbf{True} }$
			\EndIf
		\EndFor
	\EndFor
	\If { $\neg \var{overlap}$ }
		\If { $\vec{D}_\mathrm{move} \in$ cell $(x,y)$}
			\State $\set{ \vec{D}_\mathrm{sh}[i] }{ \vec{D}_\mathrm{move} }$
		\EndIf
	\EndIf
	\State $\set{i}{i+1}$
	\If {$i \ge n$}
		\State $\set{i}{0}$
	\EndIf
\EndFor
\State $\store{\vec{D}_\mathrm{sh}[i]}{\var{disk}[x,y,i]}~\forall i \in [0 \ldots n_\mathrm{max})$

\end{algorithmic}
\end{algorithm}

\begin{algorithm}[tb]
\caption{Index cell data }
\label{alg:index}
\begin{algorithmic}[1]

\Procedure{cell\_index}{x, y, i}

\If{$x \in \text{Odd}$}
\State $\set{q}{(x + m)/2}$
\Else
\State $\set{q}{x / 2}$
\EndIf

\State \Return $(i \cdot m + y) \cdot m + q$
\EndProcedure

\end{algorithmic}
\end{algorithm}

\begin{algorithm}
\caption{Cell shift GPU kernel}
\label{alg:cellshift}

\begin{algorithmic}[1]
\Require \texttt{gdim} is $(m / \var{bdim.x}, m / \var{bdim.y})$
\Require $\vec{D}_\mathrm{sh}[p]$ is stored in shared memory such that each thread indexes unique elements in memory for $p \in [0 \ldots n_\mathrm{max})$
\Procedure{shift\_cells}{$\vec{f}, d$}

\State $\set{x}{\left(\var{bidx.x}\cdot \var{bdim.x} \right)}$
\State $\set{y}{\left(\var{bidx.y}\cdot \var{bdim.y} \right)}$

\State $\load{n_\mathrm{current}}{\var{n}[x,y]}$

\State $\set{\vec{D}_\mathrm{sh}[i]}{(-10,-10)}~\forall i \in [0 \ldots n_\mathrm{max})$

\State $\set{n_\mathrm{new}}{0}$

\For {$i \in [0 \ldots n_\mathrm{current})$}
	\State $\load{ \vec{D} }{ \var{disk}[x, y,i] }$
	\State $\set{ \vec{D} }{ \vec{D} - \vec{f}\cdot d}$
	
	\If {$D.x > 0 \wedge D.y > 0 \wedge D.x \le w \wedge D.y \le w$}
		\State $\set{\vec{D}_\mathrm{sh}[n_\mathrm{new}]}{\vec{D}}$
		\State $\set{n_\mathrm{new}}{n_\mathrm{new}+1}$
	\EndIf
\EndFor

\State $\set{ (x_\mathrm{neigh}, y_\mathrm{neigh}) }{ $ cell in direction of $\vec{f}$ }
\State $\set{ \vec{s} }{ $ vector pointing to neighbor }
\State $\load{n_\mathrm{neigh}}{\var{n}[x_\mathrm{neigh},y_\mathrm{neigh}]}$

\For {$i \in [0 \ldots n_\mathrm{neigh})$}
	\State $\load{ \vec{D} }{ \var{disk}[x_\mathrm{neigh}, y_\mathrm{neigh},i] }$
	\State $\set{ \vec{D} }{ \vec{D} - \vec{f}\cdot d}$
	
	\If {$D.x > 0 \wedge D.y > 0 \wedge D.x \le w \wedge D.y \le w$}
		\State \Comment Particle stays in neighbor cell, do nothing
	\Else
		\State $\set{ \vec{D} }{ \vec{D} + \vec{s}}$
		\State $\set{\vec{D}_\mathrm{sh}[n_\mathrm{new}]}{\vec{D}}$
		\State $\set{n_\mathrm{new}}{n_\mathrm{new}+1}$
	\EndIf
\EndFor

\State $\store{\vec{D}_\mathrm{sh}[i]}{\var{disk\_dbl}[x,y,i]}~\forall i \in [0 \ldots n_\mathrm{max})$
\State $\store{ n_\mathrm{new} }{ \var{n\_dbl}[x,y] }$

\EndProcedure
\end{algorithmic}
\end{algorithm}

\autoref{alg:subsweep} implements the MPMC sub-sweep (lines 4--13 of \autoref{alg:mppmc}) update in a CUDA kernel.
One thread is launched for each cell in the active checkerboard set.
Lines 1 and 2 compute the $(x,y)$ index of the cell to which the thread is assigned.
Rows in a thread block handle rows in the cell data.
For example, threads with $x$ ids 0,1,2,3 are assigned to cells with $x$ coordinates 0,2,4,6 relative to some offset.
Similarly, each row in a thread block is assigned to alternating $y$ rows in the cell data.

Each thread initializes its own random number stream (line 3).
We use the Saru PRNG, developed by Steve Worley\cite{Worley2008}, to create uncorrelated random number streams from a hash of the thread index, current step index and a user chosen seed.
NVIDIA's CURAND library is an alternative, but requires reading and writing a large state in each thread, which slows performance by 30\%.
See Ref.~\citep{Phillips2011} for more details on the tradeoffs of various parallel PRNG schemes.

Lines 4--9 read the assigned cell into shared memory and shuffle the particles.
Line 11 starts a loop over $n_M$ trial moves.
For each selected particle $i$ in sequence, line 12 generates the trial move and lines 13--23 check for any overlaps with particles in the neighboring cells.
Lines 24-28 update the particle to its new position if the move generates no overlaps and remains in the cell.
Lines 39--32 wrap $i$ back to the start of the cell when the end is reached.

When checking overlaps, each thread reads the eight neighboring cells.
Typical GPU kernels with this memory access pattern use shared memory as a managed cache to avoid multiple reads from the same cell.
In our case, the data size per cell is large and would occupy a substantial fraction of the available shared memory, limiting parallelism.
Instead, we read only the current cell into shared memory (line 8) and use hardware cached reads for the neighbor accesses (line 18).

A 128-byte wide cache line in K20 fits four full cells in a row.
However, a row-major assignment of $[x,y,i]$ to a linear index is not ideal.
Only one particle would be read at a time from alternating cells in a row, using 8 out of the 128 bytes in a delivered cache line.
A carefully chosen mapping from $[x,y,i]$ indices to linear memory addresses leads to full utilization.
\autoref{alg:index} implements that mapping.
First, $i$ is the slowest index so that memory instructions in loops over $i$ in the kernel read contiguous data.
The next fastest index is $y$ which is handled in the traditional manner.
The $x$ index is the fastest, but it is rearranged so that all the odd, and similarly even, $x$ values are contiguous in the linear space.
We achieve further performance improvements by using texture reads ($\texttt{tex1Dfetch}$) in place of all global memory loads.
On K20, the texture cache provides the most throughput for reused read-only data.

Thanks to these efforts,  we achieve excellent utilization of the available memory bandwidth.
Benchmarks with the NVIDIA visual profiler show a sustained bandwidth of $\sim 200$ GB/s out of the texture cache.
Achieving high occupancy and limiting divergence are just as important.
For example, selecting the 16k/48k (L1/shared) mode increases occupancy and boosts performance by 50\% compared to the 48k/16k mode.
The loop on line 16 goes from 0 to $n_\mathrm{max}$ to boost performance by reducing divergent branches compared to looping over the number of particles currently in the cell.
We set the values of the empty particle slots to $(-10,-10)$ so that they do not result in false overlaps.
Early exit conditions upon finding the first overlap (not shown) cause additional divergent branches, but removing these checks reduces performance due to the increase in computations and memory accesses.

Kernel performance varies with block size~\cite{Anderson2008a}.
Short benchmarks show that (32, 4) is the fastest, and we use it for all production runs.
It outperforms the slowest by 55\%, demonstrating the importance of performing this test.

\autoref{alg:cellshift} implements the cell shift step on the GPU by equivalently translating the particles in the opposite direction.
One thread per cell gathers all of the particles that belong in the new cell and builds the list in shared memory.
It then writes out the cell to a separate memory area, $\var{disk\_dbl}$ and  $\var{n\_dbl}$, so that other running threads do not read updated data.
After the kernel completes, the double buffered data structures are swapped.
Since the shift direction is one of either $+x$, $-x$, $+y$, or $-y$, only two old cells contribute particles to the new cell: the cell with the same index, and one neighbor.
Lines 7--14 loop over the current cell, read in each particle, shift the cell, and if that particle is still within the cell boundaries it is added to the new list.
Lines 18--28 perform the same operations on the neighbor cell, with the addition of a coordinate system transformation.
These rely on the following logic: if the particle left its host cell, it must have entered this cell.
In this manner, each particle is checked for inclusion by two threads and at two separate points in the code.

The floating point operations by both of these threads \emph{must} be identical.
Consider if line 20 were to perform the coordinate system translation $\vec{D} - \vec{f}\cdot d + \vec{s}$ and then check for particles that enter the current cell.
Floating point round-off errors may result in both this check and the corresponding check on line 10 to fail, losing the particle.
In a test simulation configured with $1024^2$ particles, several hundred were lost after $10^6$ sweeps.
When implemented as shown in \autoref{alg:cellshift}, no particles are lost even after $10^9$ sweeps.

\subsection{Parameter tuning}
The maximum move radius $d$ and the number of trial moves performed in each cell update $n_M$ are free parameters.
At fixed $n_M$, we test $d \in [0.06, 0.08, \ldots 0.20]$ and at fixed $d$, we test $n_M\in[1 \ldots 8]$.
Each test measures the autocorrelation of the average orientational order parameter\cite{Anderson2012} over a long run of $10^9$ sweeps for $N=512^2$.
We find that $d=0.16$ and $n_M=4$ minimize the autocorrelation time $\tau$ when measured in wall clock seconds.


\section{Results}

The hard disk system is a standard model system in statistical mechanics and the one which was originally used to pioneer the Monte Carlo computer simulation method~\cite{Metropolis1953}.
Its phase behavior is completely determined by the equation of state, which is the relation between internal pressure $P$ and packing fraction $\phi=\rho\pi(\sigma/2)^2$.
Here, $\rho=N/V$ is density and $V$ the volume of the simulation box.
At packing fractions between $\phi=0.7$ to $0.72$ the system undergoes a first-order phase transition from the liquid phase to the hexatic phase, followed by a continuous transition to the solid phase~\cite{Bernard2011}.
In their paper, Bernard and Krauth showed with an event-chain simulation method~\cite{Bernard2009} that only very large simulations ($N>256^2$) have minimal finite size effects, and long equilibration times are necessary.
This demonstrates the need for high performance MC code and explains why previous simulation studies of the hard disk system~\cite{Alder1962, Lee1992, Zollweg1992, Weber1995, Jaster1999, Mak2006} have been unable to provide conclusive evidence for a first-order phase transition or the existence of an intermediate hexatic phase.

\begin{DIFnomarkup}

\begin{table*}
\centering
\begin{tabular}{c r l l l l l l}
\toprule
System & & \multicolumn{6}{c}{Dimensionless pressure $P^*$ in the hard disk system at packing fraction} \\
size & Method &  $\phi=0.698$ & $\phi=0.702$  & $\phi=0.706$ & $\phi=0.710$ & $\phi=0.714$ & $\phi=0.718$\\
\cmidrule(l){3-8}
$N=256^2$ & MPMC    & 9.1709(1) & 9.1920(2) & 9.1854(1) &  9.1792(1) & 9.1758(1) & 9.187(1) \\
               & BK            & 9.1708(4)   & 9.1924(4) & 9.1858(5) &  9.1790(4) & 9.1758(5) & 9.186(1) \\
               & Difference& 0.0000(5)   & 0.0004(5) & 0.0004(5) &  0.0002(4) & 0.0000(6) & 0.001(1) \\
\cmidrule(l){3-8}
$N=512^2$ & MPMC & 9.1699(5) & 9.1900(2) & 9.1861(1) & 9.1828(1) & 9.1800(1) & 9.1930(4)\\
                & BK        & 9.1700(2) & 9.1899(6) & 9.1856(6) & 9.1821(5) & 9.1803(4) & 9.1937(2)\\
                & Difference& 0.0001(5)  & 0.0001(6) & 0.0004(6) &  0.0007(5) & 0.0003(4) & 0.0006(5) \\
\cmidrule(l){3-8}
$N=1024^2$ & MPMC & 9.16934(4) & 9.1882(3) & 9.1859(3) & 9.1842(3) & 9.1819(4) & 9.1951(4)\\
                & BK                & 9.1693(1)   & 9.1880(2) & 9.1855(2) & 9.1843(2) & 9.1822(2) & 9.1949(3)\\
                & Difference& 0.0000(1)  & 0.0002(4) & 0.0003(4) &  0.0001(4) & 0.0002(5) & 0.0001(5) \\
 \bottomrule
\end{tabular}
\caption{\label{fig:eos}
This table shows data for the equation of state $P^*(\phi)$ over the range where the liquid transforms into the solid.
It includes runs by MPMC (this work) and by Bernard and Krauth (BK) with serial event chain simulation~\cite{Bernard2011}.
Error bars are shown at two standard errors of the mean, $\sigma = 2(\langle [P^* - \langle P^* \rangle]^2 \rangle / N_\mathrm{samples})^{1/2}$, where the number in parentheses is the error in the last digit shown.
Our data is averaged over 8 independent runs of $10^9$ sweeps (64 runs for $\phi=0.718$, $N=256^2$).
See Ref.\cite{Bernard2011} for a description of the BK data averaging scheme.
Differences in the pressures are shown with propagated error bars $(\sigma_\text{MPMC}^2 + \sigma_\text{BK}^2)^{1/2}$.}
\end{table*}
\end{DIFnomarkup}

Phase transitions are extremely sensitive to the slightest programming error or inadvertent correlation of trial moves due to the appearance of (quasi-)long-range spatial correlations and long equilibration times.
Structural fluctuations are more important in two dimensions than in three dimensions, and they are particularly large for the hard disk system.
These properties, together with the availability of high-precision serial data to compare with, make hard disks a good system to test our algorithm.

In principal, computing the pressure for hard disks is simple:
Estimate the radial distribution function $g(r)$ in the limit as $r$ approaches the disk diameter $\sigma$ from the right, and calculate~\cite{Metropolis1953}
\begin{equation}
P^*=\frac{P\sigma^2}{k_\mathrm{B} T} = \sigma^2\rho \left( 1 + \frac{\pi}{2} \sigma^2 \rho \lim_{r \rightarrow \sigma+} g(r)\right),
\label{eq:P}
\end{equation}
where we introduce the dimensionless pressure $P^*$.
In practice, obtaining an unbiased estimate requires special care.
We estimate pressure using the following procedure.
For each sampled configuration, a histogram of particle pair distances is computed over all $N$ particles.
$n[r_i]$ counts the number of particle pairs between $r_i$ and $r_i + \delta r$.
The pair distribution function $g(r)$ is evaluated by the equation
\begin{equation}
g(R_i) = \frac{n[r_i]}{N \rho \, \delta A} = \frac{n[r_i]}{N \rho \, 2 \pi R_i \, \delta r}
\label{eq:gR}
\end{equation}
at the sampling points $r=R_i$, where
\begin{equation}
R_i = \frac{2}{3} \frac{{r_{i+1}}^3 - {r_i}^3}{{r_{i+1}}^2 - {r_i}^2}.
\label{eq:R}
\end{equation}
The formula for $R_i$ is derived assuming a linear dependence of $n(r)$ with $r$, which we observe to be valid close to $\sigma$.
The minimum position $R_0$ is greater than $\sigma$, so direct evaluation of the limit is not possible.
We fit $g(R_i)$ to a polynomial of degree $d$ in the range $r \in (\sigma, \sigma + c]$ and extrapolate to $r=\sigma$.
Extensive testing with a model distribution ($n(r) = 30e^{-30r}$) tunes the parameters to ensure that there is no systematic bias.
We choose parameters in the middle of the flat region with systematic errors less than $10^{-5}$.
They are $\delta r = 10^{-4}\sigma$, $c = 0.02\sigma$, and $d = 5$.

\autoref{fig:eos} shows the average $P^*$ data obtained by simulations with MPMC.
Independent runs of $N=256^2$, $N=512^2$, and $N=1024^2$ particles are performed at packing fractions between $\phi=0.698$ and $\phi=0.718$, which comprises the transformation from liquid to solid.
Each run starts with a randomly generated configuration at low density and is quickly compressed to the target.
The run then continues at constant density for $10^9$ sweeps, which only takes 4 days for $N=256^2$ (15 days for $N=512^2$) to complete on a Tesla M2070 GPU.
The pressure is averaged every 200 sweeps in each run after an equilibration period of $3\cdot10^8$ sweeps.

Our data confirms the equation of state reported in Ref.~\citep{Bernard2011}. All values overlap within error bars.
We do not analyze positional order or orientation order in the dense phase emerging from the phase transition.
Such an analysis would be necessary to distinguish a hexatic phase from a solid phase and is left for a separate work~\cite{Anderson2012}.
We refer to Ref.~\citep{Anderson2012} for an in-depth comparison of the phase diagram of hard disks obtained with various algorithms, including MPMC.


\section{Performance}

\begin{table}
\centering
\begin{tabular}{ r l l}
\toprule
& K20 & E5540 \\
\cmidrule{2-3}
Hardware & custom built & HP DL2x170h\\
Mainboard & ASUS Sabertooth  & \\
Chipset & AMD 990FX & Intel 5520\\
CPU & Athlon II X4 630 & 2x Intel Xeon E5540 \\
CPU clock & 2.8 GHz & 2.53GHz \\
RAM & 16GB DDR3& 24GB DDR3 \\ 
RAM clock & 1333 MHz & 1333 MHz \\ \cmidrule{2-3}
GPU & Tesla K20 & \\
Core clock & 706 MHz & \\
Memory clock & 5200 MHz & \\
DRAM & 5 GB GDDR5 & \\ 
ECC & off & \\ \cmidrule{2-3}
OS & Gentoo & RHEL 6 \\
Architecture & x86\_64 & x86\_64 \\
CPU compiler & GCC 4.5.3 & GCC 4.7.0 \\
flags & -O3 -funroll-loops & -O3 -funroll-loops \\
GPU compiler & CUDA 5.0 & \\
flags & \emph{default} & \\
GPU driver & 310.19 \\
\bottomrule
\end{tabular}
\caption{\label{tab:config} Benchmark hardware and software configurations.}
\end{table}

We test the performance of the MPMC hard disk code using CUDA on a single Tesla K20 GPU (Kepler).
For comparison, we also implement the MPMC algorithm on the CPU and run it on our cluster nodes.
The CPU implementation uses OpenMP to parallelize across all cores on a node.
Each thread processes a single horizontal strip of the simulation domain and the innermost loop follows \autoref{alg:subsweep}.
We make no attempt to parallelize across multiple GPUs or multiple CPU nodes using MPI.
All tests are performed using single precision floating point format to store particle coordinates.
\autoref{tab:config} lists complete specifications of the hardware and software configurations of our test machines.

\subsection{Scaling with number of particles}
\begin{figure*}[htb]
\centering
\includegraphics{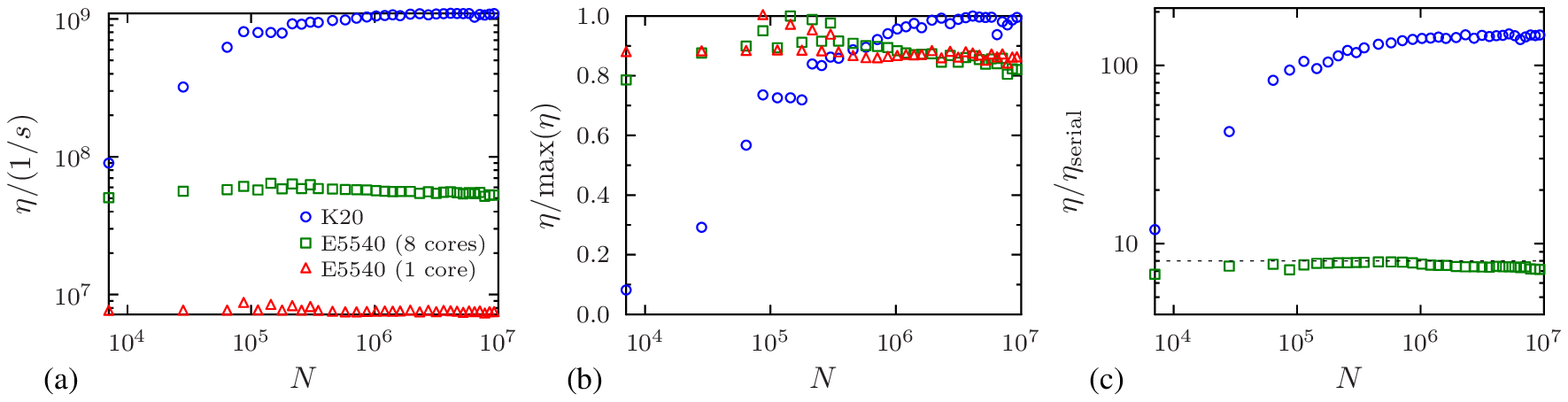}
\caption{\label{fig:bmark}
Benchmarks are performed with a varying number of disks, $N$, at a packing fraction $\phi=0.698$.
Each benchmark runs 100 sweeps to warm up and then measures the time it takes to run another 100 sweeps as well as the the number of attempted trial moves to compute $\eta = N_\mathrm{moves}/t$.
11 separate runs are performed and the median result is plotted here.
In all cases, the error bars (one standard deviation) are smaller than the symbol size.
Subfigure (a) plots the efficiency of the computation $\eta$ vs.~N on various hardware configurations.
Subfigure (b) plots $\eta$ normalized by the maximum obtained on each hardware configuration.
Subfigure (c) plots the speedup obtained over a serial execution. The dashed line marks a speedup of 8.
}
\end{figure*}

We perform benchmarks at a fixed packing fraction $\phi = 0.698$ and analyze the performance scaling with $N$.
\autoref{fig:bmark}(a) plots the results and \autoref{tab:bmark} lists selected numerical values.

\begin{table*}
\centering
\begin{tabular}{ r l l l l l}
\toprule
& & $253^2$ & $760^2$ & $1520^2$ & $3040^2$ \\
\cmidrule{3-6}

K20 & $\eta$ & $6.23 \cdot 10^8$ & $9.83 \cdot 10^8$ & $1.09 \cdot 10^9$ &  $1.09 \cdot 10^9$\\
& $\eta / \mathrm{max}(\eta)$ & 0.567  &  0.895  &  1.00 & 1.00 \\
& $\eta / \eta_\mathrm{serial}$& 83.6 & 134 & 148 & 148\\
\cmidrule{3-6}

E5540 & $\eta$  & $5.77 \cdot 10^7$ & $5.78 \cdot 10^7$  & $5.42 \cdot 10^7$ & $5.26 \cdot 10^7$ \\
(8 cores)& $\eta / \mathrm{max}(\eta)$& 0.900 & 0.900 & 0.860 & 0.820 \\
& $\eta / \eta_\mathrm{serial}$& 7.65 & 7.88 & 7.39 & 7.16 \\
\cmidrule{3-6}

E5540  & $\eta$  & $7.54 \cdot 10^6$ & $7.33 \cdot 10^6$ & $7.33 \cdot 10^6$ & $7.35 \cdot 10^6$ \\
(1 core) & $\eta / \mathrm{max}(\eta)$& 0.880 & 0.855 & 0.855 & 0.858\\
\bottomrule
\end{tabular}
\caption{\label{tab:bmark}
Numerical values for selected benchmarks from \autoref{fig:bmark}.
}
\end{table*}

The algorithm has a running time $t \in O(N)$, so the number of trial moves per unit time, $\eta$, should be constant under ideal circumstances.
In practice, there are overheads that cause deviation from constant efficiency.
\autoref{fig:bmark}(b) collapses the individual $\eta$ plots to a relative efficiency metric for each separate configuration.
When running on a single CPU core, efficiency is flat with only a slight downward drift for large $N$.
This is likely because the largest systems no longer fit in on-chip cache.
The 8-core benchmarks show the same behavior at large $N$, although it starts a factor of 2 higher because the same data is now split over 2 chips' caches.
The 8-core results also have a slight dip for $N < 10^5$ due to the overhead of managing worker threads.
On the GPU, the kernel launch overhead is significant enough that for $N$ less than $10^5$, efficiency is less than $70\%$.

Despite this inefficiency, the GPU still outperforms the single CPU core runs by a factor of 83 for systems as small as $N=253^2$.
At peak efficiency, the GPU speedup over a single core is a factor of 148.
We prefer thinking about speedups compared to a single core, but recognize that there are other ways of evaluating it.
On a per socket basis, a single GPU is 37x faster than a quad core E5540.
On a per-node basis, it is still an order of magnitude (20x) faster.
In terms of aggregate performance per price, an 8-core node with 8 externally attached GPUs (the configuration we use) is 148x faster than just the host, but only costs 5.5x as much for a benefit of 27x more sweeps/unit time given a fixed budget.
We do not have power monitoring equipment on our cluster, so we are unable to provide actual measurements of energy savings.
Instead, we obtain an estimate using the manufacturer's TDP specifications, which report 80W for the E5540 and 225W for the K20.
Based on these numbers and the per-socket speedup, a CPU simulation would use 13 times more energy than if it were run for the same number of sweeps on the GPU.

\subsection{Limitations}
The number of parallel threads depends on the cell size, interaction range, and particle shape.
In this work, MPMC executes one parallel thread per active cell.
In terms of the $N$ particles, the number of active cells is $N/(\langle n \rangle 2^d)$ where $\langle n \rangle$ is the average cell population and $d$ is the dimensionality of the system.
It is approximately $N/8$ for hard disks ($d=2$) at high density ($\langle n \rangle = 2$).
Hard spheres ($d=3$) decrease the number of active cells to $N/16$ ($\langle n \rangle = 2$).
Expanding the interaction range to a truncated and shifted Lennard-Jones potential with $r_\mathrm{cut}=2.5\sigma$ in three dimensions drops the number to $N/144$ ($\langle n \rangle = 18$ determined by MD simulation).
GPUs operate at peak efficiency only when running more than 10 thousand threads, establishing minimum practical system sizes of 80 thousand, 160 thousand, and 1.44 million particles for hard disks, hard spheres, and Lennard-Jones beads, respectively.
Depending on the desired application, these sizes may be unnecessary or prohibitively large.

Alternate thread assignment schemes for MPMC can ameliorate this problem.
One possibility is to execute $3^d$ threads per active cell, where each computes potential overlaps between the trial moves in the active cell with particles in one neighboring cell.
This technique greatly increases the number of parallel threads, reducing the theoretical minimum practical system sizes to 9 thousand, 6 thousand, and 53 thousand for hard disks, hard spheres, and Lennard-Jones beads, respectively.
These sizes are much more in line with typical simulation sizes in soft matter research.

Large density fluctuations present in Lennard-Jones systems should not pose a problem.
Thread divergence is not increased because the loop over $n_M$ trial moves in each active cell is identical across the entire system.
The GPU's fine-grained scheduler replaces blocks as soon as they finish executing, so the load is automatically balanced despite blocks in low density regions completing sooner than those in high density areas.
Memory is wasted in the sparse cell data structure, but modern GPUs have large amounts of memory (6GB), enough to store tens of millions of particles even with wasted space.
Alternate data structures could always be employed.


\section{Conclusions}
Efficient parallel algorithms are essential for the application of Monte Carlo particle simulations on current and future computer hardware.
Building on prior works utilizing checkerboard domain decomposition, we detailed an algorithm for massively parallel MC and implemented it on the GPU and the CPU.
The GPU speedup that we obtain is comparable to what has been achieved in MD simulations.
Our findings demonstrate that GPUs are well-suited for running large-scale MC simulations.
Medium scale simulations could also make good use of the GPU after some work to increase parallelism in the implementation.

Future work to parallelize on multiple GPUs using MPI will enable even larger scale simulations.
However, a straightforward implementation of the MPMC algorithm onto multiple GPUS poses one major problem.
Communication with neighboring MPI ranks is necessary after every sub-sweep update, so slow PCIe interconnect bandwidths would limit performance.
A second level of parallelization that introduces inactive regions between the MPI ranks significantly decreases communication needs.
Such a technique only requires communication at the end of a sweep when the grid shift is performed.
The number of trial moves attempted per grid cell $n_M$ could be further increased to lessen communication even more.

During the course of this work we learned that it is surprisingly difficult to implement MC in a parallel environment.
Every slight violation of the balance conditions leads to incorrect sampling, and with parallel update moves it is not simple to determine which schemes do not obey balance.
Sometimes the effect on our simulations was so small that it would be easy to miss in a typical complex practical application.
It is important to test any new parallel algorithm thoroughly in a situation where reliable results are available from serial simulations.
Hard disks are such a system.
We computed to high precision the hard disk equation of state over the density range where the liquid transforms into the solid.
Our results agree perfectly with serial event-chain Monte Carlo simulations and are compatible with the presence of a first-order liquid-hexatic phase transition~\cite{Anderson2012} -- a finding that is scientifically significant by itself.

\section{Acknowledgements}
We thank Werner Krauth and Etienne Bernard for discussions during the final stages of this work.
We also thank NVIDIA for providing the Tesla K20 used for the benchmarks in this paper.
J.A.A., M.E. and S.C.G acknowledge support by the Assistant Secretary of Defense for Research and Engineering, U.S. Department of Defense [DOD/ASD(R\&E)](N00244-09-1-0062).
Any opinions, findings, and conclusions or recommendations expressed in this publication are those of the authors and do not necessarily reflect the views of the DOD/ASD(R\&E).
E.J.\ and S.C.G.\ received support from the James S. McDonnell Foundation 21st Century Science Research Award/Studying Complex Systems, grant no.\ 220020139, and E.J.\ acknowledges support from the National Defense Science and Engineering Graduate (NDSEG) Fellowship, 32 CFR 168a.
E.J.,  T.L.G., and S.C.G.\ acknowledge support from the National Science Foundation under Award No. CHE 0624807.
Simulations were performed on a GPU cluster hosted by the University of Michigan's Center for Advanced Computing.

\bibliographystyle{model1-num-names}
\bibliography{bib-discmc}

\end{document}